\documentclass{PoS}
\usepackage{epsfig}
\usepackage{bm}
\usepackage{amssymb}
\usepackage{fontenc}
\usepackage{times}
\usepackage{mathptmx}
\usepackage{graphicx}

\title{Complex Langevin for Lattice QCD at $\bm{T=0}$ and $\bm{\mu \ge 0}$.}

\ShortTitle{Complex Langevin for Lattice QCD at $T=0$ and $\mu \ge 0$.}

\author{\speaker{D.~K.~Sinclair}%
         \thanks{This research was supported in part by US Department of Energy
         contract DE-AC02-06CH11357}
         \\
HEP Division, Argonne National Laboratory, 9700 South Cass Avenue, Argonne, 
Illinois 60439, USA\\
E-mail: \email{dks@hep.anl.gov}}

\author{J.~B.~Kogut\\
Department of Energy, Division of High Energy Physics, Washington, DC 20585,
USA\\
and\\
Department of Physics -- TQHN, University of Maryland, 82 Regents Drive, 
College Park, MD 20742, USA\\
        E-mail: \email{jbkogut@umd.edu}}

\abstract{QCD at finite quark-/baryon-number density, which describes nuclear
matter, has a sign problem which prevents direct application of standard
simulation methods based on importance sampling. When such finite density is
implemented by the introduction of a quark-number chemical potential $\mu$, this
manifests itself as a complex fermion determinant. We apply simulations using
the Complex Langevin Equation (CLE) which can be applied in such cases. However,
this is not guaranteed to give correct results, so that extensive tests are
required. In addition, gauge cooling is required to prevent runaway behaviour.
We test these methods on 2-flavour lattice QCD at zero temperature on a small 
($12^4$) lattice at an intermediate coupling $\beta=6/g^2=5.6$ and relatively 
small quark mass $m=0.025$, over a range of $\mu$ values from $0$ to 
saturation. While this appears to show the correct phase structure with a phase
transition at $\mu \approx m_N/3$ and a saturation density of $3$ at large 
$\mu$, the observables show departures from known values at small $\mu$. We 
are now running on a larger lattice ($16^4$) at weaker coupling $\beta=5.7$.
At $\mu=0$ this significantly improves agreement between measured observables
and known values, and there is some indication that this continues to small
$\mu$s. This leads one to hope that the CLE might produce correct results in
the weak-coupling -- continuum -- limit. 
}

\FullConference{34th annual International Symposium on Lattice Field Theory\\
                24-30 July 2016\\
                University of Southampton, UK}

\begin{document}

\section{Introduction}

QCD at a non-zero quark-number chemical potential $\mu$ has a complex fermion
determinant. Hence standard lattice-gauge-theory simulation methods, which are
based on importance sampling, cannot be applied directly. However, the Langevin
Equation does not rely on importance sampling, and can be adapted to complex
actions by replacing real fields by complex fields 
\cite{Parisi:1984cs,Klauder:1983nn,Klauder:1983zm,Klauder:1983sp}. 
For lattice QCD at finite $\mu$, this means promoting the $SU(3)$ gauge fields
to $SL(3,C)$.

Early attempts to simulate lattice QCD at finite $\mu$ using the Complex 
Langevin Equation (CLE) were frustrated by runaway solutions which are possible
because $SL(3,C)$ is non-compact. Recently it was realized that at least part
of the reason why this occurs is that the CLE dynamics has no resistance to
the production of unbounded fields which are unbounded gauge transformations of
bounded fields. This has led to the concept of `gauge cooling', gauge 
transforming configurations to keep them as close as possible to the $SU(3)$
manifold \cite{Seiler:2012wz}. The CLE with gauge cooling has been applied to 
QCD at finite $\mu$ at large quark mass 
\cite{Aarts:2008rr,Aarts:2013uxa,Aarts:2014bwa,Aarts:2016qrv,Langelage:2014vpa}
and with smaller quark masses on small lattices \cite{Sexty:2013ica}
and more recently to QCD at finite temperature and $\mu$ \cite{Fodor:2015doa}. 
At weak enough couplings these simulations are in agreement with results
obtained using other methods.

Even when the CLE converges to a limiting distribution, it is not guaranteed to
produce correct values for the observables unless certain conditions are
satisfied \cite{Aarts:2009uq,Aarts:2011ax,Nagata:2015uga,Nishimura:2015pba}.
The reason one needs to check the validity of the CLE for QCD is to first check
the requirement that the gauge fields evolve over a bounded region, which 
appears to be true. Secondly, the CLE can only be shown to converge to the
correct distribution if the `drift terms' -- the derivatives of the (effective)
action with respect to the fields -- are holomorphic functions of the fields.
Because the fermion determinant has zeros, the drift term is only meromorphic
in the fields. Hence the CLE will only give correct results if the
contribution of the poles in the drift term are negligible. Those of the above
mentioned papers, which perform CLE simulations of QCD at finite $\mu$,
provide tests of the range of validity of the method.

Recent work reported by Aarts \cite{aarts} and by Stamatescu \cite{stamatescu}
presents methods of determining when poles in the drift term of the CLE are
likely to produce incorrect results. Studies using random-matrix theory
indicate the range of validity of the CLE and suggest modifications of gauge
cooling which can extend this range \cite{Nagata:2016alq,Zafeiropoulos}. There
is also recent work which suggests other criteria for determining when the CLE
will produce correct results and when it will fail \cite{Nagata:2016vkn}.
Other studies indicate how the introduction of irrelevant terms to the drift
term can direct the CLE to converge to correct limiting distributions
\cite{jaeger}.

We simulate lattice QCD at zero temperature and finite $\mu$ on a $12^4$ 
lattice at $\beta=6/g^2=5.6$ and $m=0.025$. For these parameters the expected
position of the transition from hadronic to nuclear matter at 
$\mu \approx m_N/3 \approx 0.33$ is well separated from any false transition
at $\mu \approx m_\pi/2 \approx 0.21$. We observe that our results are 
consistent with a transition at $\mu \approx m_N/3$, but not with the 
expectation that observables will be fixed at their $\mu=0$ values for
$\mu < m_N/3$. At large enough $\mu$ the quark number density does saturate at
$3$ as expected. Very preliminary results of these simulations were reported at
Lattice 2015 \cite{Sinclair:2015kva}.

We are now simulating on a $16^4$ lattice at weaker coupling, $\beta=5.7$, and
$m=0.025$. At $\mu=0$ we find that the observables are in far better agreement
with known results than for $\beta=5.6$. We are now moving to $\mu > 0$. We
see preliminary indications that for small $\mu$, the observables are still in
better agreement with known results than was true at $\beta=5.6$. This leads us
to hope that the CLE will converge to the correct distributions in the 
continuum -- weak coupling -- limit.

\section{Complex Langevin Equation for finite density Lattice QCD}

If $S(U)$ is the gauge action after integrating out the quark fields, the
Langevin equation for the evolution of the gauge fields $U$ in Langevin 
time $t$ is:
\begin{equation}
-i \left(\frac{d}{dt}U_l\right)U_l^{-1} = -i \frac{\delta}{\delta U_l}S(U)
+\eta_l
\end{equation}
where $l$ labels the links of the lattice, and 
$\eta_l=\eta^a_l\lambda^a$. Here $\lambda_a$ are the Gell-Mann 
matrices for $SU(3)$. $\eta^a_l(t)$ are Gaussian-distributed random 
numbers normalized so that:
\begin{equation}
\langle\eta^a_l(t)\eta^b_{l'}(t')\rangle=\delta^{ab}\delta_{ll'}\delta(t-t')
\end{equation}

The complex-Langevin equation has the same form except that the $U$s are now
in $SL(3,C)$. $S$, now $S(U,\mu)$ is 
\begin{equation}
S(U,\mu) = \beta\sum_{_\Box} \left\{1-\frac{1}{6}{\rm Tr}[UUUU+(UUUU)^{-1}]
\right\} - \frac{N_f}{4}{\rm Tr}\{\ln[M(U,\mu)]\}
\end{equation}
where $M(U,\mu)$ is the staggered Dirac operator. Note: backward links
are represented by $U^{-1}$ not $U^\dag$. Note also that we have 
chosen to keep the noise-vector $\eta$ real. $\eta$ is gauge-covariant under
$SU(3)$, but not under $SL(3,C)$. This means that gauge-cooling is non-trivial.
Reference~\cite{Nagata:2015uga} indicates why this is not expected to change
the physics. After taking $-i\delta S(U,\mu)/\delta U_l$, the cyclic
properties of the trace are used to rearrange the fermion term so that it
remains real for $\mu=0$ even after replacing the trace by a stochastic
estimator.

To simulate the time evolution of the gauge fields we use the partial 
second-order formalism of Fukugita, Oyanagi and Ukawa.
\cite{Ukawa:1985hr,Fukugita:1986tg,Fukugita:1988qs} 

After each update, we gauge-fix iteratively to a gauge which minimizes the 
unitarity norm -- gauge cooling \cite{Seiler:2012wz}:
\begin{equation}
F(U) = \frac{1}{4V}\sum_l{\rm Tr}\left[U_l^\dag U_l + (U_l^\dag U_l)^{-1} 
     - 2\right] \ge 0,
\end{equation}
where $V$ is the space-time volume of the lattice.

\section{Zero temperature simulations on a $\bm{12^4}$ lattice}

We simulate lattice QCD with 2 flavours of staggered quarks at finite $\mu$ on
a $12^4$ lattice with $\beta=5.6$ and quark mass $m=0.025$, using the CLE with
gauge cooling. $\mu$ is in the range $0 \le \mu \le 1.5$ which includes the
expected phase transition at $\mu \approx m_N/3 \approx 0.33$ and that of the
phase-quenched theory at $\mu \approx m_\pi/2 \approx 0.21$. ($m_N$ and $m_\pi$
are from the HEMCGC collaboration \cite{Bitar:1990cb,Bitar:1993rk,Bitar:1990wk}
). The upper limit $\mu=1.5$ lies well within the saturation regime where each
lattice site is occupied by one quark of each colour.

We simulate for 1--3 million updates of the gauge fields at each $\mu$ value.
The input updating increment $dt=0.01$. Since we use adaptive rescaling of $dt$
to control the size of the drift term, the actual $dt$s used in the updates are
considerably smaller than this. The length of the equilibrated part of the run
at each $\beta$ then lies in the range 100--1000 langevin time units. We record
the plaquette (action), the chiral condensate and the quark-number density 
every 100 updates, and the unitarity norm after each update.

\begin{figure}[htb]
\parbox{2.9in}{
\epsfxsize=2.9in
\epsffile{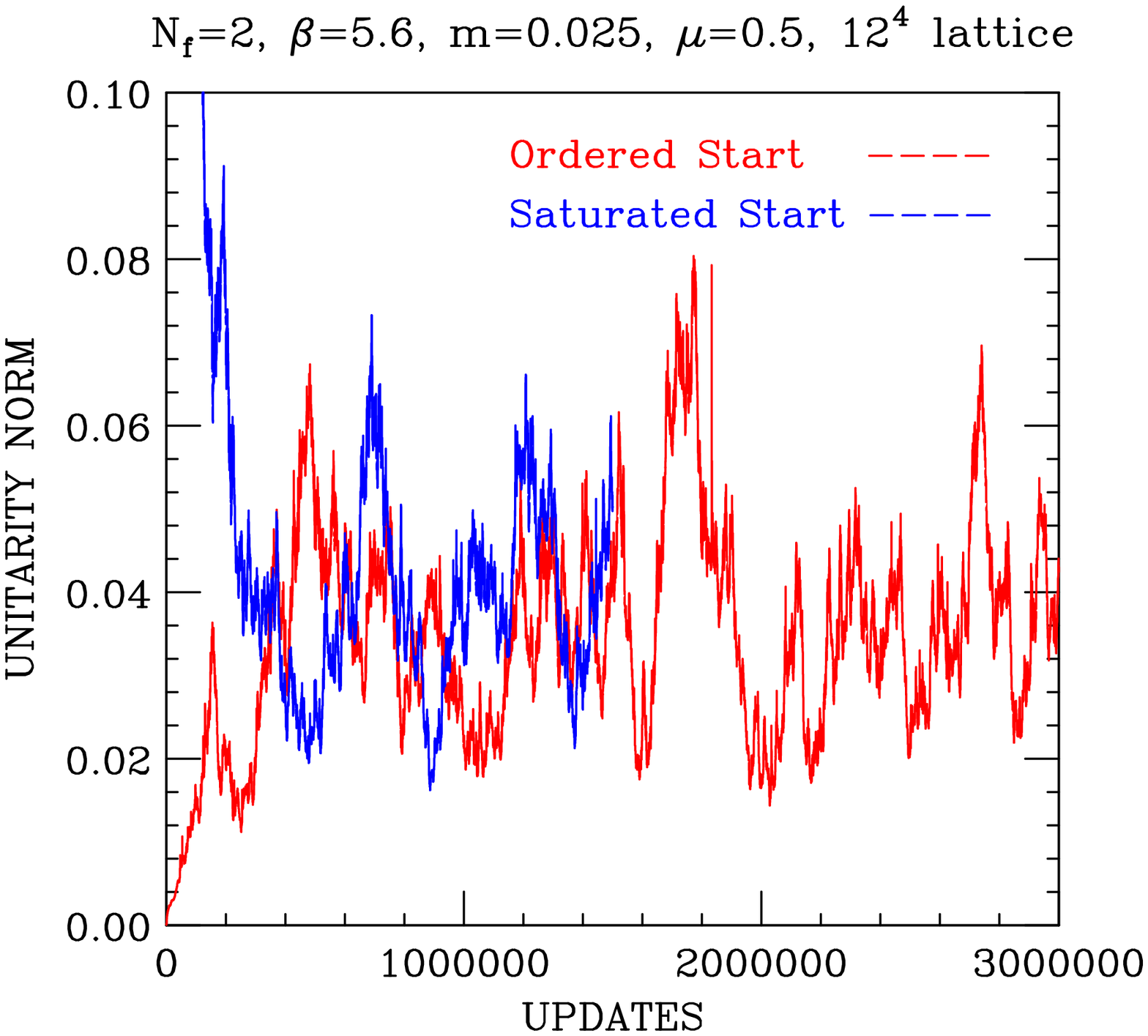}
\caption{Unitarity norms for $\mu=0.5$ on a $12^4$ lattice. The red curve is
for the run starting from an ordered start. The blue curve is for the run
starting from a $\mu=1.5$ configuration.}
\label{fig:unorm5}
}
\parbox{0.2in}{}
\parbox{2.9in}{
\epsfxsize=2.9in 
\epsffile{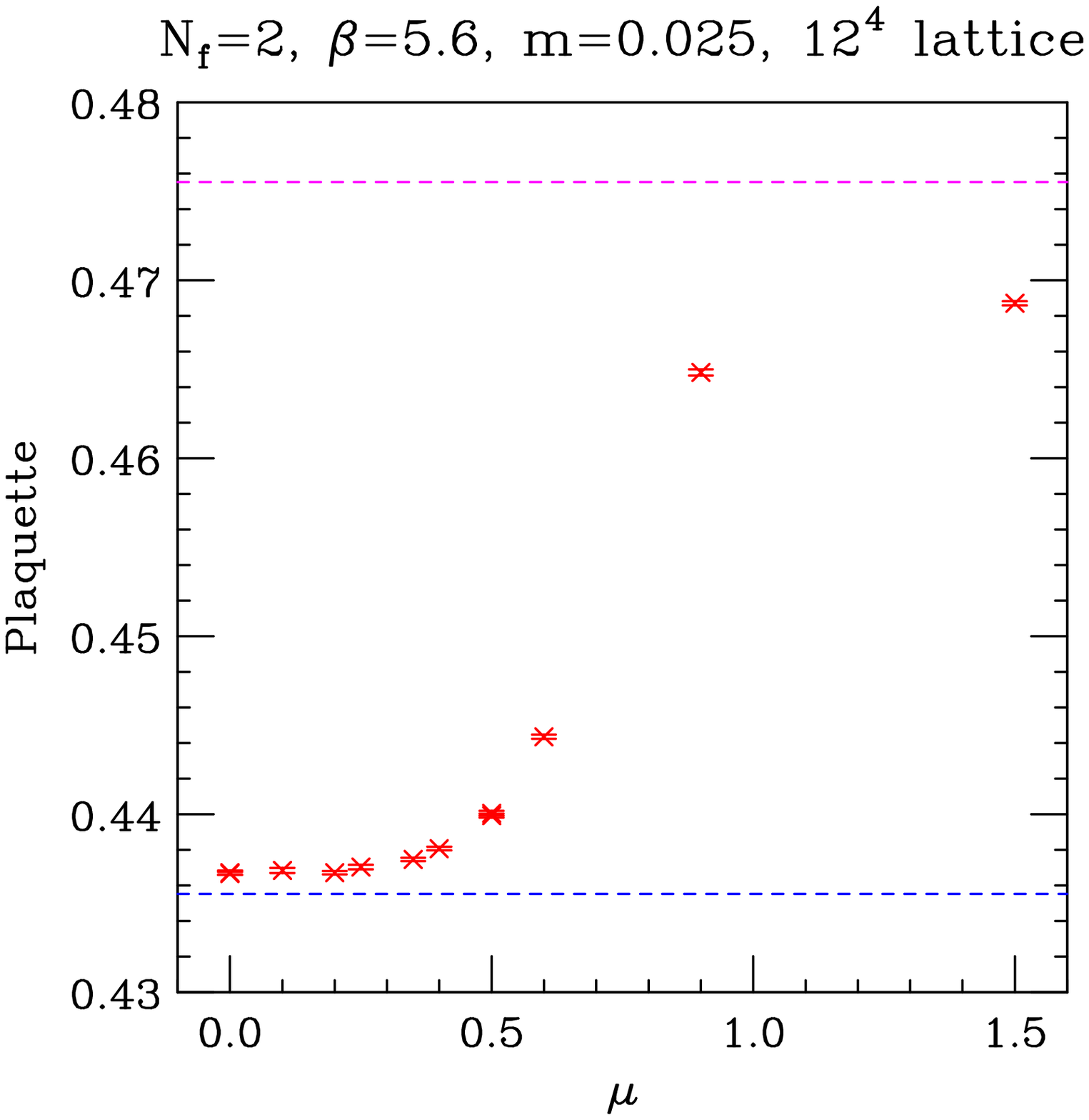}
\caption{Plaquette as a function of $\mu$. Dashed lines are the correct value
at $\mu=0$ and the quenched value.}
\label{fig:apq}
}
\end{figure}

\begin{figure}[hbt]
\parbox{2.9in}{
\epsfxsize=2.9in
\centerline{\epsffile{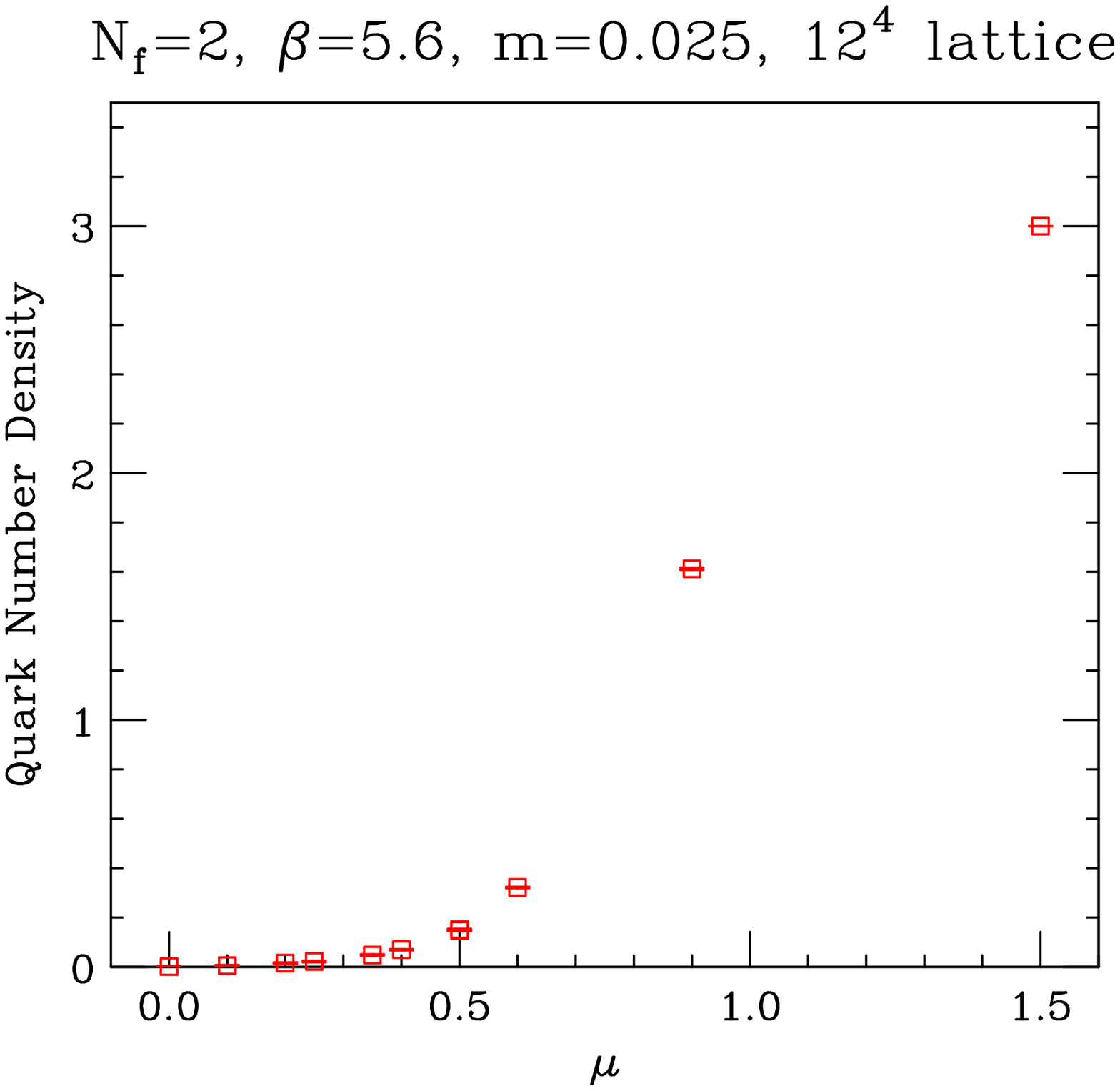}}
\caption{Quark number density, normalized to one staggered quark (4-flavours), 
as a function of $\mu$.}
\label{fig:qnd}
}
\parbox{0.2in}{}
\parbox{2.9in}{
\epsfxsize=2.9in
\centerline{\epsffile{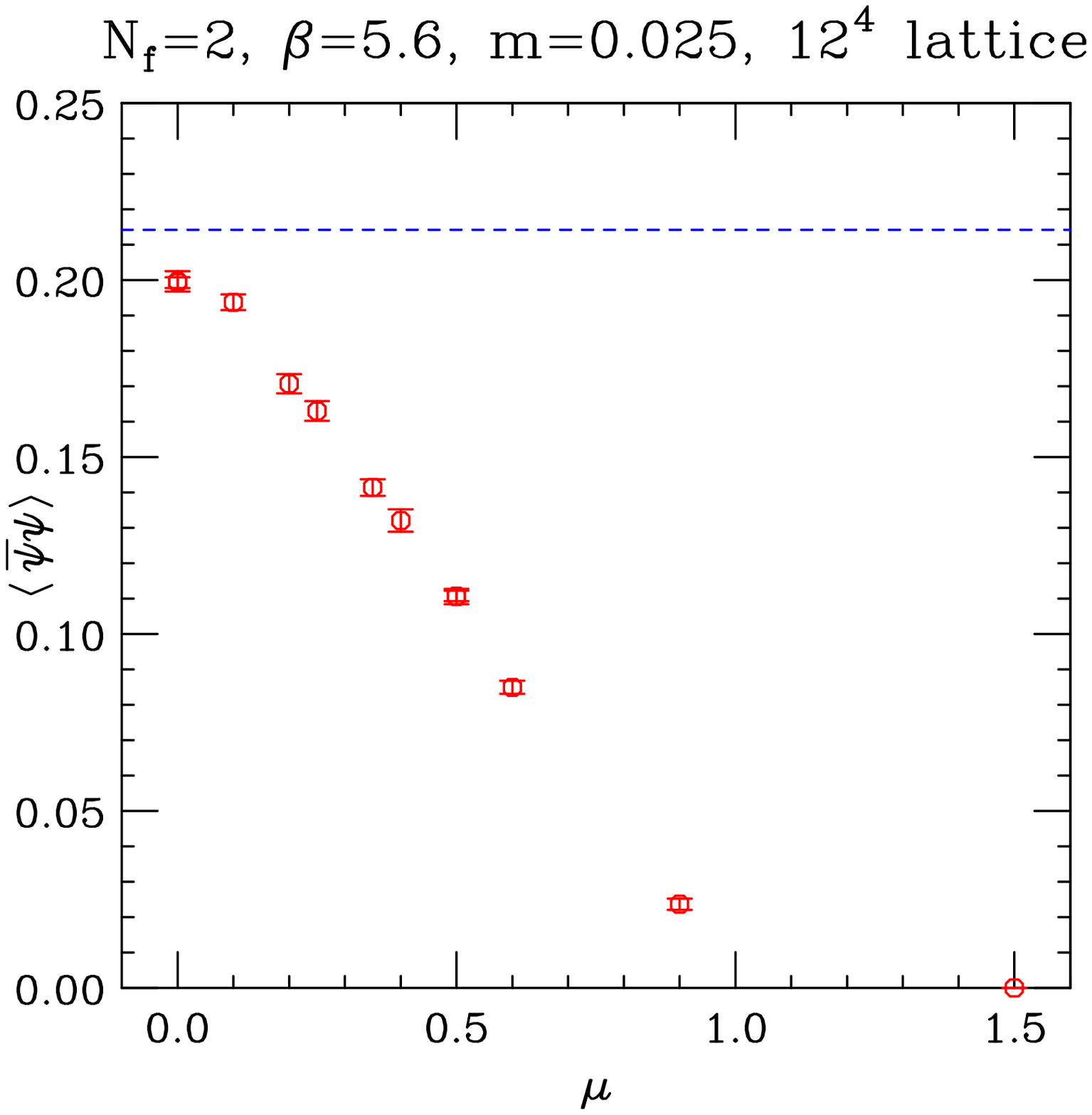}}
\caption{Chiral condensate, normalized to one staggered quark (4-flavours), as 
a function of $\mu$. Dashed line is the correct value at $\mu=0$.}
\label{fig:pbp} 
}                                                             
\end{figure}    

At each $\mu$ we observe that the unitarity norm appears to evolve over a
compact domain, which is one of the requirements for the CLE observables to
have a well-defined limit. It is also a necessary but not sufficient condition
for it to produce correct results. At $\mu=0$ and $\mu=0.5$ we have produced
trajectories from both an ordered start and starting from an equilibrated
configuration at $\mu=1.5$. In both cases, it appears that the compact domain
is independent of the start, as are the average observables. 
Figure~\ref{fig:unorm5} shows the evolution of the unitary norms at $\mu=0.5$
from the 2 different starts. It is interesting to note that the unitarity norm
has a minimum somewhere in the range $0.35 \le \mu \le 0.9$. Does this mean
that the CLE produces correct results for $\mu$ sufficiently large?

Figure~\ref{fig:apq} shows the plaquette as a function of $\mu$ from these runs.
We note that there is a very small but significant difference between the
value at $\mu=0$ and the correct value obtained from an RHMC simulation. The
real Langevin equation yields a value significantly closer to the correct value,
so this deviation is not due solely to the inexact nature of the Langevin 
method. For $\mu \le 0.25$, the plaquette appears to be (almost) independent of
$\mu$ as expected. For $\mu \ge 0.35$ the plaquette increases with $\mu$
up until saturation.

Figure~\ref{fig:qnd} shows the quark-number density as a function of $\mu$.
For $\mu \le 0.25$ this number density is small -- it is expected to be zero.
For $\mu \ge 0.35$ this number density increases, reaching the saturation 
value of 3 (3 quarks of different colours at each site), for large $\mu$.
We note, however, that this density does not appear to show an abrupt increase
at the transition as might be expected for a first-order phase transition.
 
In figure~\ref{fig:pbp} we plot the chiral condensate 
($\langle\bar{\psi}\psi\rangle$) as a function of $\mu$. At $\mu=0$ it already
lies appreciably below the exact value. Instead of remaining constant up to the
phase transition to nuclear matter as expected , it starts to fall 
monotonically once $\mu > 0$, finally reaching the expected value of zero at 
saturation.

Hence for $\beta=5.6$, $m=0.025$ on a $12^4$ lattice, the CLE appears to 
produce the correct phase structure, although the phase transition at 
$\mu \approx m_N/3$ does not show any evidence for its expected first-order
behaviour. The plaquette shows small deviations from the correct values for
small $\mu$ as does the quark-number density. The chiral condensate shows
larger departures from its expected behaviour.

\section{Zero temperature simulations on a $\bm{16^4}$ lattice}

We are now running CLE simulations on a $16^4$ lattice. At $\beta=5.6$, 
$m=0.025$, comparison with our $12^4$ runs indicates that finite size effects
are small as are finite $dt$ errors.

This larger lattice allows us to run at weaker coupling. We are now running at
$\beta=5.7$, $m=0.025$. For our $\beta=5.6$, $m=0.025$ runs at $\mu=0$, the
CLE measured plaquette value is $0.43690(6)$ compared with the RHMC value
$0.43552(2)$, while the chiral condensate is $0.1974(7)$ compared with
$0.2142(8)$ for the RHMC. At $\beta=5.7$, $m=0.025$, the CLE measured plaquette
value is $0.42374(4)$ compared with the RHMC value $0.42305(1)$, so the
systematic error has been reduced by roughly a factor of 2. For the chiral
condensate the CLE value is $0.1738(11)$ compared with the RHMC value of
$0.1754(2)$, almost an order of magnitude improvement. This gives us some hope
that the CLE will give correct values for observables in the weak-coupling
(continuum) limit. We are now extending these $\beta=5.7$, $m=0.025$ simulations
to non-zero $\mu$.

\section{Discussion, Conclusions and Future Directions}

We simulate 2-flavour Lattice QCD at finite $\mu$ on a $12^4$ lattice at 
$\beta=5.6$, and light quark mass $m=0.025$ using the CLE with gauge cooling.
We see indications of the expected phase transition from hadronic to nuclear
matter at $\mu \approx m_N/3$, and the passage to saturation at large $\mu$.
There are, however, systematic departures from known and expected results.
At $\mu=0$ the plaquette and chiral condensates disagree with known
results. For the plaquette the systematic error is very small and for 
$\mu < m_N/3$ the plaquette is almost independent of $\mu$ as expected.
At small $\mu$, the chiral condensate decreases with increasing
$\mu$ rather than remaining constant. These do not appear to be a 
finite-size effects. The reason for these systematic errors is presumably 
because zeros of the fermion determinant produce poles in the drift term, which
prevent it from being holomorphic in the fields, a requirement for proving the
validity of the CLE. These zeros also produce poles in the chiral condensate,
which could explain why it shows larger departures from expected values than
do other observables.

We are extending our simulations to $16^4$ lattices. In addition to showing
that finite size (and finite $dt$) effects are small, these allow us to
simulate at smaller coupling, $\beta=5.7$. Here, simulations at $\mu=0$ show
that systematic errors are significantly reduced. This leads to the hope that,
in the weak coupling (continuum) limit, the CLE might yield correct results
(after continuing to $dt=0$). Preliminary results from simulations with 
$\mu > 0$ look promising.

Modifications to the CLE designed to reduce failures of the method need to be
pursued. These include modifications to gauge cooling \cite{Nagata:2016alq},
and modifications to the dynamics by the introduction of irrelevant operators
either to the action or to the drift term directly \cite{jaeger}.

We plan to extend our zero-temperature simulations to smaller quark masses.
Finite temperature simulations are also planned.

Once it is known that the CLE is generating correct results, we will study
the high-$\mu$ phase for signs of colour superconductivity. This will also
require simulations for $N_f=3$ and $N_f=2+1$. At finite temperature we will
search for the critical endpoint.

\section*{Acknowledgements}
These simulations are performed on Edison and Cori at NERSC, Comet at SDSC,
Bridges at PSC, Blues at LCRC Argonne and Linux PCs in HEP Argonne.

\end{document}